\documentstyle[prd,aps]{revtex}

\begin{document}

\title{Second order perturbations of a Schwarzschild black hole:
inclusion of  odd parity perturbations}

\author{Carlos O. Nicasio$^1$, Reinaldo Gleiser$^1$, Jorge Pullin$^2$ }

\address{1. Facultad de Matem\'atica, Astronom\'{\i}a y F\'{\i}sica,
Universidad Nacional de C\'ordoba,\\ Ciudad
Universitaria, 5000 C\'ordoba, Argentina.}
\address{2. Center for Gravitational Physics and Geometry, Department of
Physics,\\
The Pennsylvania State University,
104 Davey Lab, University Park, PA 16802}
\maketitle

\begin{abstract}
We consider perturbations of a Schwarzschild black hole that can be of 
both even and odd parity, keeping terms up to second order in perturbation
theory, for the $\ell=2$ axisymmetric case. We develop explicit
formulae for the evolution equations and radiated energies and
waveforms using the Regge--Wheeler--Zerilli approach. This formulation
is useful, for instance, for the treatment in the ``close limit
approximation'' of the collision of counterrotating black holes.
\end{abstract}


\section{Introduction}

There has been a recent surge of interest in black hole perturbation
theory, stemming from the successful applicability of such a formalism
in the case of black hole collisions (see \cite{PuKy} for a
review). In particular, the use of second order perturbative
computations \cite{second} has proven quite useful to endow the
perturbative formalism with ``error bars'' and also to increase its
accuracy.  

The computations associated with the second order formulation are
quite laborious. In particular, the second order perturbative
equations are linear differential equations that contain ``source''
terms that are quadratic in the first order perturbations. This
forbids us from doing a ``generic'' second order computation, since it
would involve an infinite number of source terms. Computations have to
be restricted to certain multipolar orders and parities to achieve a
computable expression. The expressions we have presented in the past
\cite{second} are all for even parity axisymmetric $\ell=2$ first
order perturbations. This is a significant case, since for first order
perturbations such modes are the ones that dominate the initial data
for black hole collisions (at least in the head-on case, in the
non-head on case one also has non-axisymmetric $m=\pm 2$ modes). If
the black holes individually have spin, however, odd parity terms
appear as well. If one wishes to study such collisions with the second
order formulation one needs to lay out the second order equations for
odd and even parity perturbations. Since the ``source'' terms that
appear in the second order equations are quadratic, one has quadratic
contributions from odd parity first order modes in the even parity
second order equations. The intention of this paper is to present a
comprehensive account of the perturbative equations involving odd
terms, up to second order. In a separate publication we will apply the
formalism to the collision of counterrotating black holes.

\section{Perturbative formalism: review of some first order results}

\subsection{The Regge--Wheeler decomposition and the Regge--Wheeler gauge}
In order to fix notation and to set the computational philosophy  for
the rest of the paper, we recollect here several results of first
order perturbation theory that are known, but that it is useful to
have laid out in a compact fashion for even and odd perturbations
together. 

We consider the axially symmetric perturbations of a spherically
symmetric background, up to and including second order, using the
Regge--Wheeler framework.  In first order perturbations there is no
loss of generality in the consideration of axisymmetric perturbations
since modes with different $m$ value decouple. This is not the case,
however, in second order theory and our treatment is therefore 
restricted to the axisymmetric case. In terms of the usual Schwarzschild-like
coordinates $t,r,\theta, \phi$, any $\ell=2$  axially symmetric
perturbation may be written in the ``Regge--Wheeler form'' \cite{ReWh}
 \begin{eqnarray}
\widetilde{g}_{tt}&=&-(1-2M/r) \left\{1 - \left[\epsilon
\widetilde{H}^{(1)}_0(r,t)+ \epsilon^2
\widetilde{H}^{(2)}_0(r,t)\right] P_2(\theta) \right\}\label{rw1}\\
\widetilde{g}_{rr}&=&(1-2M/r)^{-1} \left\{1 +\left[\epsilon
\widetilde{H}^{(1)}_2(r,t)+\epsilon^2 \widetilde{H}^{(2)}_2(r,t)
\right] P_2(\theta)\right\}\\ \widetilde{g}_{rt}&=&\left[\epsilon
\widetilde{H}^{(1)}_1(r,t)+\epsilon^2
\widetilde{H}^{(2)}_1(r,t)\right] P_2(\theta)\\ \widetilde{g}_{t
\theta}&=& \left[\epsilon \widetilde{h}^{(1)}_0(r,t)+ \epsilon^2
\widetilde{h}^{(2)}_0(r,t)\right]\partial P_2(\theta) /\partial
\theta)\\
\widetilde{g}_{t \phi}&=& \left[\epsilon \widetilde{h}^{(1)}_{0o}(r,t)+ 
\epsilon^2 \widetilde{h}^{(2)}_{0o}(r,t)\right] \sin(\theta) \partial
P_2(\theta) /\partial \theta)\\
\widetilde{g}_{r \phi}&=& \left[\epsilon \widetilde{h}^{(1)}_{1o}(r,t)+ 
\epsilon^2 \widetilde{h}^{(2)}_{1o}(r,t)\right] \sin(\theta) \partial
P_2(\theta) /\partial \theta)\\
\widetilde{g}_{r \theta}&=& \left[ \epsilon
\widetilde{h}^{(1)}_1(r,t)+ \epsilon^2
\widetilde{h}^{(2)}_1(r,t)\right]\partial P_2(\theta) /\partial
\theta)\\ 
\widetilde{g}_{\theta\theta}&=& r^2 \left\{1+[\epsilon
\widetilde{K}^{(1)}(r,t)+\epsilon^2 \widetilde{K}^{(2)}(r,t)]
P_2(\theta) \right. \nonumber \\ & & \left.+ [\epsilon
\widetilde{G}^{(1)}(r,t) + \epsilon^2 \widetilde{G}^{(2)}(r,t)]
\partial^2 P_2(\theta) /\partial \theta^2)\right\}\\
\widetilde{g}_{\phi\phi} &=&r^2 \left\{\sin^2\theta+ \sin^2\theta [\epsilon
\widetilde{K}^{(1)}(r,t) + \epsilon^2 \widetilde{K}^{(2)}(r,t)]
P_2(\theta) \right. \nonumber \\ & & \left.+\sin(\theta) \cos(\theta)[
\epsilon \widetilde{G}^{(1)}(r,t) +\epsilon^2
\widetilde{G}^{(2)}(r,t)] \partial P_2(\theta) /\partial
\theta)\right\}\\
\widetilde{g}_{\theta \phi}&=& {1 \over 2} \left[\epsilon
\widetilde{h}^{(1)}_{2o}(r,t)+ \epsilon^2
\widetilde{h}^{(2)}_{2o}(r,t)\right] \left[ \cos(\theta) \partial
P_2(\theta)/\partial\theta- \sin(\theta) \partial^2
P_2(\theta)/\partial\theta^2 \right] \label{rwe}
\end{eqnarray}
where $P_2(\theta)=3 \cos^2(\theta)/2-1/2$. We introduce the expansion
parameter $\epsilon$, which defines the perturbation order, and
identify the corresponding metric coefficients with the superscripts
$(1)$, $(2)$ and use the subscript ``o'' to refer to the odd parity 
quantities.

As we stated above, any axisymmetric perturbation can be put into the
Regge--Wheeler form. However, for calculational simplicity, it is
usually convenient to take advantage of the coordinate freedom to 
restrict somewhat the form of the metric. An example of such choice
is the so called  {\em Regge - Wheeler gauge} (RWG), 
where the nonvanishing metric
coefficients are
\begin{eqnarray}
g_{tt}&=&-(1-2M/r) \left\{1 -[\epsilon H^{(1)}_0(r,t)+\epsilon^2
H^{(2)}_0(r,t)] P_2(\theta) \right\}\\ g_{rr}&=&(1-2M/r)^{-1} \left\{1
+[\epsilon H^{(1)}_2(r,t) +\epsilon^2 H^{(2)}_2(r,t)]P_2(\theta)
\right\}\\ g_{rt}&=&[\epsilon H^{(1)}_1(r,t) +\epsilon^2
H^{(2)}_1(r,t)] P_2(\theta)\\ g_{t\theta}&=&[\epsilon
h^{(1)}_{0o}(r,t) +\epsilon^2 h^{(2)}_{0o}(r,t)] \sin(\theta)\partial
P_2(\theta)/\partial\theta\\ g_{r\theta}&=&[\epsilon h^{(1)}_{1o}(r,t)
+\epsilon^2 h^{(2)}_{1o}(r,t)] \sin(\theta)\partial
P_2(\theta)/\partial\theta\\ g_{\theta\theta}&=& r^2
\left\{1+[\epsilon K^{(1)}(r,t) +\epsilon^2 K^{(2)}(r,t)] P_2(\theta)
\right\}\\ g_{\phi\phi} &=&r^2 \left\{\sin^2\theta+ \sin^2\theta [\epsilon
K^{(1)}(r,t)+ \epsilon^2 K^{(2)}(r,t)] P_2(\theta) \right\},
\end{eqnarray}
that is,  we are demanding that $h^{(i)}_0 = h^{(i)}_1 = G^{(i)} =
h^{(i)}_{2o} = 0$, $i=1,2$.

A remarkable aspect of the Regge--Wheeler gauge (RWG) is that it is {\em
unique}, in the sense that the metric perturbation coefficients in the RWG 
can be uniquely recovered from those in an arbitrary gauge. That is, given an
arbitrary metric, in an arbitrary gauge, there is a well defined
procedure that brings it to the RWG.

Let us exhibit this property for first order gauge transformations. 
One can uniquely choose a gauge transformation vector that brings 
the metric to the RWG form \cite{second}. The resulting
transformation formulae are,
\begin{eqnarray}
K^{(1)} & = & \widetilde{K}^{(1)} +(r-2M) \left
( \widetilde{G}^{(1)},_r -{2 \over r^2}\widetilde{h}^{(1)}_1 \right)
\\ H^{(1)}_2 & = & \widetilde{H}^{(1)}_2 + (2r-3M)
\left(\widetilde{G}^{(1)},_r - {2 \over r^2} \widetilde{h}^{(1)}_1
\right) + r(r-2M) \left(\widetilde{G}^{(1)},_r - {2 \over r^2}
\widetilde{h}^{(1)}_1 \right)_{,r} \\ H^{(1)}_1 & = &
\widetilde{H}^{(1)}_1+r^2 \widetilde{G}^{(1)},_{tr} -
\widetilde{h}^{(1)}_1,_t - {2M \over r(r-2M)} \widetilde{h}^{(1)}_0 +
\widetilde{h}^{(1)}_0,_r + {r(r-3M) \over r-2M} \widetilde{G}^{(1)},_t
\\ H^{(1)}_0 & = & \widetilde{H}^{(1)}_0 -M
\left(\widetilde{G}^{(1)},_r - {2 \over r^2 } \widetilde{h}^{(1)}_1
\right) - {2 r \over r-2M } \widetilde{h}^{(1)}_0,_t +{r^3 \over
(r-2M)} \widetilde{G}^{(1)},_{tt}\\ h^{(1)}_{0o} & = & {1 \over 2}
\widetilde{h}^{(1)}_{2o,t} + \widetilde{h}^{(1)}_{0o}\\ h^{(1)}_{1o} &
= & {1 \over {2 r}} \left( r \widetilde{h}^{(1)}_{2o,r} - 2
\widetilde{h}^{(1)}_{2o} \right) + \widetilde{h}^{(1)}_{1o}.
\end{eqnarray}

Something that should be stressed is that given the uniqueness of the
RWG, quantities computed in such a gauge (like, for instance, metric 
components), are {\em gauge invariant}. That is, substituting the 
above formulae in the definition of any quantity in the RWG one
obtains expressions for the quantity in any gauge.

The procedure can in principle be repeated at every order in
perturbation theory. To achieve this, one first performs a first order
transformation that yields the metric (at first order) in the
RWG. This transformation will induce transformations at all higher
orders as well. One then makes a {\em purely second order} gauge
transformation that brings the second order portion of the metric to
the RWG. This will leave the first order piece in RWG, and will modify
the third and higher orders as well. One can continue this procedure
up to an arbitrarily high order and the metric will be, up to that
order, in the RWG. Unfortunately, the procedure is not quite unique at
higher orders.  The reason for this is that if before performing the
second order gauge transformation, one performs an $\ell=0$ first
order gauge transformation, the procedure we outlined yields at the
end of the day a {\em different} second order metric, but {\em still
in the RWG}.  That is, the metric one obtains at the end of the
procedure is not unique, in spite of being in the RWG. Fortunately,
one can isolate second order quantities that are invariant under these
$\ell=0$ transformations.  This point has been discussed in reference
\cite{boost2} in detail, so we will not repeat it here. From the
expressions for waveforms given in this paper, one can apply the
techniques of reference \cite{boost2} in order to compare with
numerical results without the $\ell=0$ ambiguity.

From now on, we will do computations in the RWG. Because of what we
just discussed, there is no issue of gauge invariance in our
formalism, since all quantities in the RWG (appropriately modified to
take into account the $\ell=0$ issue) will be ``gauge invariant'' in
the sense described above.

\subsection{The first order Zerilli equation}

As shown by Zerilli, we
may introduce a function $\psi^{(1)}(r,t)$, such that if we write
for the metric components in the RWG,
\begin{eqnarray}
\label{K1}
K^{(1)}(r,t) & = & 6 {r^2+rM +M^2 \over
r^2(2r+3M)}\psi^{(1)}(r,t)+\left(1-2 {M
\over r} \right) {\partial \psi^{(1)}(r,t)\over \partial r}   \\
\label{H0}
H^{(1)}_0(r,t) & = & {\partial \over \partial r}\left[{2r^2-6rM-3M^2
\over
r(2r+3M)} \psi^{(1)}(r,t)+(r-2M) {\partial \psi^{(1)}(r,t) \over
\partial r}
\right] - K^{(1)}(r,t)   \\
\label{H1}
H^{(1)}_1(r,t) & = & {2r^2-6rM-3M^2 \over (r-2M)(2r+3M)} {\partial
\psi^{(1)}(r,t)
\over \partial t}+r {\partial^2 \psi^{(1)}(r,t) \over \partial r
\partial t}
 \\
H^{(1)}_2(r,t) & = & H^{(1)}_0(r,t) \label{H12},
\end{eqnarray}
then the linearized Einstein equations are satisfied, provided only that
$\psi^{(1)}(r,t)$ is a solution of the Zerilli equation (\ref{Zer21}).
\begin{equation}
{\partial^2 \psi^{(1)}(r,t) \over \partial {r^*}^2} - {\partial^2
\psi^{(1)}(r,t)
\over \partial t^2} -V_Z(r^*) \psi^{(1)}(r,t) =0
\label{Zer21}
\end{equation}
where
\begin{equation}
r^*=r+2M \ln[r/(2M)-1]
\end{equation}
and
\begin{equation}
V_Z(r) = 6\left(1-2{M \over r}\right){4 r^3+4 r^2 M+6 r M^2+3 M^3
\over r^3 (2 r+3 M)^2 }
\end{equation}

It is straightforward to obtain ``inversion" formulas for
$\psi^{(1)}(r,t)$ in terms of the metric coefficients. From (\ref{K1})
and (\ref{H1}), we have
\begin{equation}
{\partial \psi^{(1)}(r,t) \over \partial t} = {r \over 2r+3M} \left[ r
{\partial K^{(1)} \over \partial t} - \left(1 -2 {M \over r} \right)
H^{(1)}_1 \right]
\label{psi1a}
\end{equation}
while from (\ref{K1}) and (\ref{H0}) we find
\begin{equation}
\psi^{(1)}(r,t) = {r(r-2M) \over 3(2r+3M)} \left( H^{(1)}_0 - r
{\partial K^{(1)} \over \partial r} \right) + { r \over 3} K^{(1)}
\label{psi1b}
\end{equation}

Equations (\ref{psi1a}) and (\ref{psi1b}) are equivalent, as far as
first order perturbations are concerned, up to an additive function of
$r$.  This ambiguity is of no concern since the relevant physical
quantities, like radiated energies and waveforms are all computed in
terms of the time derivative of $\psi^{(1)}$, so it is conventionally
ignored, we will assume we set the additive function to zero. Equation
(\ref{psi1b}) can be used to obtain a gauge invariant form for
$\psi^{(1)}(r,t)$, i.e., an expression for $\psi^{(1)}(r,t)$, valid in
an {\em arbitrary gauge}. This is an especially important result,
because it allows us to obtain the initial data for the Zerilli
equation, directly in an appropriate gauge. It should be noted,
however, that although $\psi^{(1)}(r,t)$ is uniquely defined in a
general gauge by (\ref{psi1b}), we do not have, in a general gauge,
unique expressions for the metric perturbation functions in terms of
$\psi^{(1)}(r,t)$. The general statement of Zerilli's results is, in
this case, that if the metric perturbations satisfy the first order
Einstein equations, then $\psi^{(1)}(r,t)$ satisfies the Zerilli
equation (\ref{Zer21}).

The situation, as we shall see, is somewhat different for the second
order perturbations, because of the presence of ``source terms" in the
Zerilli equation.

\subsection{The Odd parity equation}

One can also write a ``Zerilli"-like equation (known as the
Regge--Wheeler equation, historically it was discovered earlier than
the even parity one \cite{ReWh}) for the odd part of first order metric
perturbations and the equations they satisfy. If we define
\begin{eqnarray}
h^{(1)}_{0o}(r,t) & = & \left(1- {2M\over r}\right) \left( r{\frac {\partial }
{\partial r}}Q^{(1)}(r,t)+Q^{(1)}(r,t)\right)\\ h^{(1)}_{1o}(r,t) & =
& \frac{r}{1-2M/r} {\frac {\partial }{\partial t}}Q^{(1)}(r,t)
\end{eqnarray}
the Einstein equations are satisfied, provided that
$Q^{(1)}(r,t)$ is a solution of the equation
\begin{equation}
{\partial^2 Q^{(1)}(r,t) \over \partial {r^*}^2} - {\partial^2
Q^{(1)}(r,t)
\over \partial t^2} -V_Q(r^*) Q^{(1)}(r,t) =0
\label{eqQ}
\end{equation}
where
\begin{equation}
r^*=r+2M \ln[r/(2M)-1]
\end{equation}
and
\begin{equation}
V_Q(r) = 6 \frac{(r-M)(r-2M)}{r^4}.
\end{equation}

It is easier to obtain inversion formulas in this case
that in the even parity case.  We can write,
\begin{eqnarray}
Q^{(1)}(r,t) & = & - {1 \over 4} \left( r {\partial h^{(1)}_{1o}(r,t)
\over \partial t} - r {\partial h^{(1)}_{0o}(r,t) \over \partial r} +
2 h^{(1)}_{0o}(r,t) \right) \label{Q0}\\ { \partial Q^{(1)}(r,t) \over
\partial t} & = & \frac{(1-2 M/r)}{r} h^{(1)}_{1o}(r,t) \label{Qt0}
\end{eqnarray}
We can make the same comments as in the even case. Equation (\ref{Q0})
can be used to give a gauge invariant form for the odd parity Zerilli
function. Both equations allow us to give initial conditions for
$Q^{(1)}(r,t)$ in an any gauge. As in the even case, there is no unique
way to write the metric perturbations in terms of $Q^{(1)}(r,t)$ in an
arbitrary gauge.

\section{Second order perturbations}

We define the {\em second order Regge--Wheeler gauge} by imposing that
$h^{(i)}_0 = h^{(i)}_1 = G^{(i)} = h^{(i)}_{2o} = 0$, $i=1,2$. It is
easy to check that this gauge is uniquely defined, as long as we
restrict to $\ell =2$ axisymmetric first and second order gauge
transformations, but the general expressions for the second order gauge
transformations are rather lengthy.

It should be clear from the perturbation expansion, that the second
order Einstein equations imply, in turn, that the second order
perturbation functions satisfy the same linear set of equations as the
first order perturbations, but with additional terms quadratic in the
first order perturbations, that may be thought of as representing
``sources" for these equations.

Therefore if motivated by the corresponding expression for the time
derivatives $ \partial \psi^{(1)} / \partial t$, $\partial Q^{(1)} /
\partial t$, of the Zerilli functions in first order perturbation
theory, we introduce gauge invariant functions, given by
\begin{eqnarray}
\label{chi1}
\chi^{(2)}(r,t) & = & {r \over 2r+3M} \left[ r {\partial K^{(2)} (r,t)\over
\partial t}
-\left(1 - {2M \over r} \right) H^{(2)}_1(r,t) \right]\\
\Theta^{(2)}(r,t) & = & \frac{(1-2 M/r)}{r} h^{(2)}_{1o}(r,t) \label{Q2}
\end{eqnarray}
it follows that if the second order Einstein equations are satisfied, then $
\chi^{(2)}(r,t)$, $\Theta^{(2)}$ satisfy equations of the form

\begin{eqnarray}
{\partial^2 \chi^{(2)}(r,t) \over \partial {r^*}^2} - {\partial^2
\chi^{(2)}(r,t)
\over \partial t^2} -V_Z(r^*) \chi^{(2)}(r,t) +{\cal{S}}_Z =0
\label{chi2}\\
{\partial^2 \Theta^{(2)}(r,t) \over \partial {r^*}^2} - {\partial^2
\Theta^{(2)}(r,t)
\over \partial t^2} -V_Q(r^*) \Theta^{(2)}(r,t) +{\cal{S}}_Q =0
\end{eqnarray}
where ${\cal{S}}_Z$, ${\cal{S}}_Q$ are quadratic polynomials in the
first order functions and their $t$ and $r$ derivatives, with $r$
dependent coefficients. ${\cal{S}}_Z$,${\cal{S}}_Q$ may be considered
as a kind of "source terms" for the otherwise homogeneous equations
satisfied by $\chi^{(2)}$, $\Theta^{(2)}$.

But, precisely because of the presence of these ``source terms", if we
redefine $\chi^{(2)}(r,t)$, $\Theta^{(2)}(r,t)$ by the addition of a
quadratic polynomial in the first order functions and their
derivatives, the new, redefined $\chi^{(2)}(r,t)$, $\Theta^{(2)}(r,t)$ will
still satisfy an equation of the form (\ref{chi2}), provided only that
the first and second order Einstein equations are satisfied.

This arbitrariness in the definition of the Zerilli functions
associated to the second order perturbations may be used to simplify
the form, and asymptotic behavior of the ``source terms" $
{\cal{S}}_Z$, $ {\cal{S}}_Q$. As was discussed in \cite{second}, the
source terms are delicate in the sense that they can be divergent for large
values of $t$ and $r$. The divergence of the source does not imply a
physical divergence in the problem, but it poses serious difficulties
for numeric computations.  If we make the following choice, based on
the study of the solutions of the Zerilli equations for large values
of $r$, the source simplifies and also is well behaved at spatial
infinity,
\begin{equation}
\chi^{(2)}(r,t)  = {r \over 2r+3M} \left[ r {\partial K^{(2)} \over
\partial t} - \left(1-{2M \over r} \right) H^{(2)}_1 \right] -{2 \over
7} \left[ {r^2 \over 2r+3M} K^{(1)} {\partial K^{(1)} \over \partial
t} + ( K^{(1)} )^2 \right]
\label{chi2r}
\end{equation}
as our definition for $\chi^{(2)}(r,t)$, while $\Theta^{(2)}(r,t)$ is
defined by (\ref{Q2}). Making the
appropriate replacements, we then find for the sources,
\begin{eqnarray}
{\cal{S}}_Z & = & 
-{12\over 7}{(r-2M)^{3} \over r^3 (2 r+3 M)} \left[
{\frac  {\left (-24 {r}^{5}-108 M{r}^{4}-72
 {M}^{2}{r}^{3}+24 {M}^{3}{r}^{2}+180 {M}^{4}r+180 {M}^{5}\right )
\left ( \psi \right )^{2}}{{r}^{5}\left (2 r+3 M\right )\left (r-2
 M\right )^{2}}} \right. \nonumber \\ 
 &+&{\frac  {\left (64 {r}^{5}+176 M{r}^{4}+592 {M}^{2
}{r}^{3}+1020 {M}^{3}{r}^{2}+1122 {M}^{4}r+540 {M}^{5}\right )
 \psi_r  \psi }{{r}^{4}\left (2 r+3 M\right )^{2}\left (r-2 M
\right )}}\nonumber \\ 
 &-&{\frac  {\left (112 {r}^{5}+480 M{r}^{4}+692 {M}^{2}{r}^{
3}+762 {M}^{3}{r}^{2}+441 {M}^{4}r+144 {M}^{5}\right ) \psi_t 
 \psi }{{r}^{2}\left (2 r+3 M\right )^{3}\left (r-2 M\right )^{2}}
}\nonumber \\ 
 &+&{\frac  {\left (8 {r}^{3}+16 {r}^{2}M+36 r{M}^{2}+24 {M}^{3}
\right ) \psi_{rr}  \psi }{{r}^{3}\left (2 r+3 M\right )}}
+{\frac  {
\left (8 {r}^{2}+12 rM+7 {M}^{2}\right ) \psi_{rt}  \psi }{r\left (
2 r+3 M\right )\left (-r+2 M\right )}}
+{\frac  {M \psi  \psi_{rrt} }{2 r+3 M}}
+{\frac {r \psi_t  \psi_{rrr} }{3}} \nonumber \\ 
 &-&{\frac  {\left (12 {r}^{3}+36 {r}^{2}M+59 r{M}^{2}+90 
{M}^{3}\right )\left ( \psi_r \right )^{2}}{3 {r}^{3}\left (r-2 M
\right )}}
 +{\frac  {\left (-18 {r}^{3}+4 {r}^{2}M+33 r{M}^{2}+48 {M
}^{3}\right ) \psi_r  \psi_t }{3 \left (2 r+3 M\right )r\left (r-2
 M\right )^{2}}}\nonumber \\ 
 &-&{\frac  {\left (12 {r}^{2}+20 rM+24 {M}^{2}\right 
) \psi_r  \psi_{rr} }{3 {r}^{2}}}
-{\frac  {\left (-3 r+7 M\right )
 \psi_r  \psi_{rt} }{-3 r+6 M}}
-{\frac  {r \psi_r  \psi_{rrt} }{3}} \nonumber \\
&+&{\frac {12 \left ({r}^{2}+rM+{M}^{2}\right )^{2}\left ( \psi_t \right 
)^{2}}{\left (2 r+3 M\right )r\left (r-2 M\right )^{3}}}
-{\frac  {
\left (-2 {r}^{2}+{M}^{2}\right ) \psi_{rr}  \psi_t }{\left (2 r+3 
M\right )\left (r-2 M\right )}}
+{\frac  {\left (4 {r}^{2}+4 rM+4 {
M}^{2}\right ) \psi_t  \psi_{rt} }{\left (r-2 M\right )^{2}}} \nonumber\\
 &-&{\frac  {\left (2 r+3 M\right )
\left (r-2 M\right )\left ( \psi_{rr} \right )^{2}}{3 r}}
 -{\frac  {
\left (2 r+3 M\right )r\left ( \psi_{rt} \right )^{2}}{-3 r+6 M}}-
{\frac {\left (-2 r+10 M\right )\left (3 r+8 M\right ) Q_r  Q_t }{
r\left (r-2 M\right )^{2}}}\nonumber \\ 
 &+&{\frac  {\left (22 {r}^{2}+70 rM+66 {M
}^{2}\right ) Q_r  Q_{rt} }{\left (2 r+3 M\right )\left (r-2 M
\right )}}
- {\frac {\left (-136 {r}^{3}-462
 {r}^{2}M+50 r{M}^{2}+456 {M}^{3}\right ) Q  Q_t }{{r}^{2}\left (2
 r+3 M\right )\left (r-2 M\right )^{2}}}\nonumber  \\ 
 &+& \left. {\frac  {\left (14 {r}^{2
}+58 rM+66 {M}^{2}\right ) Q  Q_{rt} }{r\left (2 r+3 M\right )
\left (r-2 M\right )}}
-{\frac  {\left (22 r+54 M\right )\left (r+M\right ) Q_t 
 Q_{rr} }{\left (2 r+3 M\right )\left (r-2 M\right )}} \right]
\end{eqnarray}

\begin{eqnarray}
{\cal{S}}_Q & = &
-{12 \over 7}  {(r-2 M)^{3} \over r^3 (2 r+3 M)} \left[
{\frac {\left (40 {r}^{4}-150 M{
r}^{3}-138 {r}^{2}{M}^{2}+162 {M}^{3}r+81 {M}^{4}\right ) Q 
 \psi_t }{3 {r}^{3}\left (2 r+3 M\right )\left (r-2 M\right )^{2}
}}\nonumber \right. \\ 
 &+&{\frac {\left (2 r+3 M\right )\left (-r+M\right ) Q  \psi_{rrt} }
{r\left (r-2 M\right )}}
+{\frac {\left (4 {r}^{3}-17 {r}^{2}M-33 
r{M}^{2}+81 {M}^{3}\right ) Q  \psi_{rt} }{3 {r}^{2}\left (r-2 M
\right )^{2}}}\nonumber \\ 
 &+&{\frac {\left (-160 {M}^{2}{r}^{3}+216 {M}^{5}+102 {
M}^{3}{r}^{2}-160 M{r}^{4}+387 {M}^{4}r-128 {r}^{5}\right ) Q_t 
 \psi }{2 {r}^{3}\left (2 r+3 M\right )^{2}\left (r-2 M\right )^{
2}}}\nonumber \\ 
 &+&{\frac {\left (75 {M}^{3}{r}^{2}-12 {r}^{5}+81 {M}^{5}-10 {M}
^{2}{r}^{3}+14 M{r}^{4}+150 {M}^{4}r\right ) \psi  Q_{rt} }{{r}^{2}
\left (2 r+3 M\right )^{2}\left (r-2 M\right )^{2}}}
-{\frac {\left (2 r+3 M\right )r \psi_{rr}  Q_{rrt} }{3}}\nonumber \\ 
 &+&{\frac {
\left (4 {r}^{3}+4 {r}^{2}M+6 r{M}^{2}+3 {M}^{3}\right ) \psi 
 Q_{rrt} }{r\left (2 r+3 M\right )\left (r-2 M\right )}} 
 -{\frac {
\left (-8 {r}^{4}+56 M{r}^{3}-40 {r}^{2}{M}^{2}-39 {M}^{3}r+72 {M
}^{4}\right ) Q_t  \psi_r }{2 {r}^{2}\left (2 r+3 M\right )\left (-
r+2 M\right )^{2}}}\nonumber \\ 
 &-&{\frac {\left (-18 {r}^{4}+19 M{r}^{3}-13 {r}^
{2}{M}^{2}+78 {M}^{3}r+81 {M}^{4}\right ) \psi_r  Q_{rt} }{3 r
\left (2 r+3 M\right )\left (r-2 M\right )^{2}}}
+{\frac {\left (2
 {r}^{2}-2 rM+3 {M}^{2}\right ) \psi_r  Q_{rrt} }{-3 r+6 M}}\nonumber \\
&-&{\frac {\left (28 {r}^{4}-72 M{r}^{3}-96 {r}^{2}{M}^{2}+114 {M}^{3}
r+81 {M}^{4}\right ) \psi_t  Q_r }{3 {r}^{2}\left (2 r+3 M\right )
\left (r-2 M\right )^{2}}}
+{\frac {\left (-2
 {r}^{2}+6 rM+3 {M}^{2}\right ) \psi_t  Q_{rrr} }{-3 r+6 M}}\nonumber \\
 &-&{\frac {\left (2 {r}^{4}-9 M{r}^{3}+9 
{r}^{2}{M}^{2}+22 {M}^{3}r+9 {M}^{4}\right ) \psi_t  Q_{rr} }{r
\left (2 r+3 M\right )\left (r-2 M\right )^{2}}}
+{\frac {\left (4 {r}^{3}-11 {r}^{2}M-15 r{M}^{2}+63 {M}^{3}\right )
 Q_t  \psi_{rr} }{3 r\left (r-2 M\right )^{2}}}\nonumber \\ 
 &+&{\frac {\left (6 {
r}^{2}+7 rM+18 {M}^{2}\right ) \psi_{rr}  Q_{rt} }{-6 r+12 M}}
-{\frac {
\left (10 {r}^{3}-10 {r}^{2}M-37 r{M}^{2}+27 {M}^{3}\right )
 \psi_{rt}  Q_r }{3 r\left (r-2 M\right )^{2}}}\nonumber \\ 
 &-&{\frac {\left (6 {
r}^{2}+7 rM+18 {M}^{2}\right ) \psi_{rt}  Q_{rr} }{-6 r+12 M}}
+{\frac {\left (2 r+3 M\right )r Q_{rrr}  \psi_{rt} }{3}}
-{\frac {
\left (2 r+3 M\right )\left (-r+3 M\right ) Q_t  \psi_{rrr} }{-3 r
+6 M}}\nonumber \\ 
 &-&{\frac {\left (2 r+3 M\right )r \psi_{rrr}  Q_{rt} }{3}}
+{
\frac {\left (2 r+3 M\right )\left (-r+M\right ) \psi_{rrt}  Q_r }{-
3 r+6 M}}
+ \left. {\frac {\left (2 r+3 M\right )r Q_{rr}  \psi_{rrt} }{3}}
\right]
\end{eqnarray}
where $\psi = \psi^{(1)}$, $Q = Q^{(1)}$ and the subindices indicate 
partial derivatives.

\section{Gravitational wave amplitude}

A straightforward computational way to define a gravitational wave
amplitude, is to consider a gauge in which the space-time is
manifestly asymptotically flat. For instance, consider the typical
gauge discussed in reference \cite{MTW}. In this gauge the metric has
components that behave in the following way,
\begin{eqnarray}
H_2 & \simeq & O(1/r^3) \nonumber \\
h_1 & \simeq & O(1/r) \nonumber \\
K   & \simeq & O(1/r) \nonumber \\
G   & \simeq & O(1/r) \nonumber \\
h_{1o} & \simeq & O(1/r) \nonumber \\
h_{2o} & \simeq & O(r) \label{mfar}
\end{eqnarray}
and, of course, $H_0 = H_1 = h_0 = h_{0o} = 0$. The gravitational wave
amplitude can then simply be read off from the components $K$, $G$ and
$h_{2o}$. We therefore need to relate the behavior of these components
of the metric to the Zerilli function in an asymptotic region far away
from the source.

In order to study the asymptotic behavior, one considers a solution of
the Zerilli equation in terms of an expansion in inverse powers of $r$
with coefficients that are functions of the retarded time $t-r^*$, as
discussed in detail in \cite{second}. Substituting in the Zerilli
equations we obtain,
\begin{eqnarray}
\psi^{(1)} & = & F_{\psi}^{(iii)}(t-r^*)+{3\over r}
F_{\psi}^{(ii)}(t-r^*) -{3\over 4 r^2} \left[ -7
F_{\psi}^{(ii)}(t-r^*) M+4F_{\psi}^{(i)}(t-r^*) \right] +O(1/r^3)
\nonumber \\ Q^{(1)} & = & F_Q^{(iii)}(t-r^*)+{3\over r}
F_Q^{(ii)}(t-r^*) \nonumber \\ & & +{1\over r^2}\left[-{3 \over 2} M
F_Q^{(ii)}(t-r^*)+3 F_Q^{(i)}(t-r^*) \right] +O(1/r^3) \label{azer1}
\end{eqnarray}

For this behavior of first order Zerilli functions, the
sources $S_Z$, $S_Q$ for second order Zerilli functions $\chi^{(2)}$,
$\Theta^{(2)}$ decay as $O(1/r^2)$ asymptotically. Therefore,
$\chi^{(2)}$,$\Theta^{(2)}$ will behave in a similar way to $\psi^{(1)}$,
$Q^{(1)}$ for large $r$:

\begin{eqnarray}
\chi^{(2)} & = & F^{(20)}_\chi(t-r^*)+{1\over
r}F^{(21)}_\chi(t-r^*)+O(1/r^2) \nonumber \\ \Theta^{(2)} & = &
F^{(20)}_Q(t-r^*)+{1\over r} F^{(21)}_Q(t-r^*)+O(1/r^2) \label{azer2}
\end{eqnarray}

We can now compute the asymptotic form for the metric perturbations,
to first order and second order, in Regge--Wheeler gauge, using
(\ref{azer1}) and (\ref{azer2}) in (\ref{H0}) and
(\ref{H12}). Unfortunately, in the Regge--Wheeler gauge the metric is
not in the manifestly asymptotically flat form (\ref{mfar}). We therefore
need to perform a gauge transformation to bring it to such a form, to
first and second order.

In order to do this, we start by recalling that for any $\ell=2$
axisymmetric gauge transformation, we can write the gauge
transformation vectors as,
\begin{equation}
\begin{array}{rclrcl}
\xi^{(1)t}_{(even)} & = & {\cal M}^{(1)}_0(r,t) P_2(\cos(\theta))  &
 \xi^{(1)r}_{(even)} & = & {\cal M}^{(1)}_1(r,t) P_2(\cos(\theta)) \\
\xi^{(1)\theta}_{(even)} & = & {\cal M}^{(1)}(r,t) 
\partial P_2(\cos(\theta)) / \partial \theta &
 \xi^{(1)\phi}_{(even)} & = & 0
\end{array}
\end{equation}
for even parity, and
\begin{equation}
\begin{array}{rclrcl}
\xi^{(1)t}_{(odd)} & = & 0 &
\xi^{(1)r}_{(odd)} & = & 0 \\
\xi^{(1)\theta}_{(odd)} & = & 0 &
\xi^{(1)\phi}_{(odd)} & = & {\cal M}^{(1)}_2(r,t) \sin(\theta)^{-1}
\partial P_2(\cos(\theta)) / \partial \theta
\end{array}
\end{equation}
for the odd parity case.

To achieve the asymptotically flat gauge (\ref{mfar}) we demand that 
$H^{(1)}_0(r,t)= H^{(1)}_1(r,t)= h^{(1)}_0(r,t)= h^{(1)}_{0o}= 0$. 
{}From the general form of the gauge transformation equations we have,
\begin{eqnarray}
 0 & = & \widetilde{H}^{(1)}_0 + {2M \over r(r-2M)} {\cal M}^{(1)}_1
+2 {\partial {\cal M}^{(1)}_0 \over \partial t} \nonumber \\ 0 & = &
\widetilde{H}^{(1)}_1 - {r \over r-2M} {\partial {\cal M}^{(1)}_1\over
\partial t} +{r-2M \over r} {\partial {\cal M}^{(1)}_0\over \partial
r} \nonumber \\ 0 & = & - r^2 {\partial {\cal M}^{(1)}\over \partial
t} +{r-2M \over r} {\cal M}^{(1)}_0 \nonumber \\
\label{gauge2}
{H}^{(1)}_2 & = & \widetilde{H}^{(1)}_2+ {2M \over r(r-2M)} {\cal
M}^{(1)}_1 -2 {\partial {\cal M}^{(1)}_1\over \partial r } \\
{h}^{(1)}_1 & = & - {r \over r-2M} {\cal M}^{(1)}_1 - r^2 {\partial
{\cal M}^{(1)}\over \partial r} \nonumber \\ {G}^{(1)} & = & - 2 {\cal
M}^{(1)} \nonumber \\ {K}^{(1)} & = & \widetilde{K}^{(1)} - {2 \over
r} {\cal M}^{(1)}_1 \nonumber\\ 0 & = & \widetilde{h}^{(1)}_{0o} + r^2
{\partial {\cal M}^{(1)}_2(r,t) \over \partial t} \nonumber\\ h^{(1)}_{1o} & =
& \widetilde{h}^{(1)}_{1o} + r^2 {\partial {\cal M}^{(1)}_2(r,t) \over
\partial r} \nonumber\\ h^{(1)}_{2o} & = & - 2 r^2 {\cal
M}^{(1)}_2(r,t)\nonumber
\end{eqnarray}
where the quantities with tildes are given in the Regge--Wheeler gauge,
$\{{\cal M}^{(1)},{\cal M}^{(1)}_0,{\cal M}^{(1)}_1\}$, 
are the components of the even parity gauge transformation vector, and 
${\cal M}^{(1)}_2$ that of the odd parity gauge transformation vector.

We can see that, given $\widetilde{H}^{(1)}_0$ and
$\widetilde{H}^{(1)}_1$, the first and second equation above may be
used to solve for ${\cal M}^{(1)}_0$ and ${\cal M}^{(1)}_1$. Then, the
third equation leads to a partial differential equation for 
${\cal M}^{(1)}$ while the equation for
$h^{(1)}_{0o}$ fixes ${\cal M}^{(1)}_2$. Therefore, the set of
equations, and therefore the gauge choice, is consistent. The solution,
assuming only that $\widetilde{H}^{(1)}_0$, $\widetilde{H}^{(1)}_1$,
$\widetilde{H}^{(1)}_2$, and $\widetilde{K}^{(1)}$ are given, is non
unique since one has to  integrate partial differential equations, 
so the solution in general involves free functions. 
One may choose such functions in
a judicious way to ensure that the non-vanishing  metric components 
behave in a manner consistent with asymptotic flatness.

Similarly, we define the second order asymptotically flat gauge by
 imposing the first order asymptotically flat  gauge conditions and
 $H^{(2)}_0(r,t)=0$, $ H^{(2)}_1(r,t)=0$, $ h^{(2)}_0(r,t)=0$ and $
 h^{(2)}_{0o}(r,t)=0$.  Clearly, the same reasoning as above, applied
 to the second order gauge transformations, shows that this is a
 consistent choice.

The end result of these calculations is that one may write  the first 
order gauge transformation vectors in the form,
\begin{eqnarray}
{\cal M}^{(1)}(r,t) & = & {1\over 2 r} F_\psi^{(iii)}(t-r^*)+{1\over
r^2} F_\psi^{(ii)}(t-r^*) +{1\over r^3} \left[ {1\over 4}M
F_\psi^{(ii)}(t-r^*)+{3\over 2}F_\psi^{(i)}(t-r^*)\right]+O(1/r^4)
\nonumber \\
{\cal M}^{(1)}_0(r,t) & = & {1\over 2} r F_\psi^{(iv)}(t-r^*) + \left[
F_\psi^{(iii)}(t-r^*) + M F_\psi^{(iv)}(t-r^*) \right] \nonumber \\ &
& +{1\over r}\left[ {9\over 4} M F_\psi^{(iii)}(t-r^*) +{3\over 2}
F_\psi^{(ii)}(t-r^*) + 2 M^2 F_\psi^{(iv)}(t-r^*) \right] + O(1/r^2)
\nonumber \\
{\cal M}^{(1)}_1(r,t) & = & {1\over 2} r F_\psi^{(iv)}(t-r^*)+{3\over
2} F_\psi^{(iii)}(t-r^*) \nonumber \\ & & +{1\over r}\left[ {3\over 2}
F_\psi^{(ii)}(t-r^*)-{3\over 4} M
F_\psi^{(iii)}(t-r^*)\right]+O(1/r^2) \nonumber \\
{\cal M}^{(1)}_2(r,t) & = & {3\over r} F_Q^{(iii)}(t-r^*)+{2\over r^2}
F_Q^{(ii)}(t-r^*) \nonumber \\ & & +{1\over r^3}\left[ 3
F_Q^{(i)}(t-r^*) + {1\over 2} M F_Q^{(ii)}(t-r^*) \right]+O(1/r^4)
\end{eqnarray}
The second order gauge equations are more complex then the first order
ones. It can be seen that the equations can also be solved iteratively
for all orders in $r$, but for simplicity we only compute the terms
that are going to be relevant in the gravitational waveforms.
The meaningful terms are given by
\begin{eqnarray}
{\cal M}^{(2)} & = & {1\over r}{\cal M}^{(20)}(t-r^*) + O(1/r^2)
\nonumber \\ {\cal M}^{(2)}_0 & = & r {\cal M}^{(20)}_0(t-r^*) +
O(r^0) \nonumber \\ {\cal M}^{(2)}_1 & = & r {\cal M}^{(20)}_1(t-r^*)
+ {\cal M}^{(21)}_1(t-r^*) + O(1/r) \nonumber \\ {\cal M}^{(2)}_2 & =
& {1\over r}{\cal M}^{(20)}_2(t-r^*)+ O(1/r^3)
\end{eqnarray}
and if we compute them in terms of $F_\psi$,$F_Q$, we finally get
\begin{eqnarray}
{\partial {\cal M}^{(20)}(t-r^*)\over \partial t} &=& {1\over 2}
F_\chi^{(20)}(t-r^*) - {1\over 28} F_\psi^{(v)}(t-r^*)
F_\psi^{(iii)}(t-r^*)- {1\over 28} F_\psi^{(iv)}(t-r^*)^2 - {3\over
14} F_Q^{(iv)}(t-r^*)^2 \nonumber \\
{\cal M}^{(20)}_0(t-r^*) &=& {1\over 2} F_\chi^{(20)}(t-r^*)
- {1\over 14} F_\psi^{(v)}(t-r^*) F_\psi^{(iii)}(t-r^*)
- {3\over 14} F_Q^{(iv)}(t-r^*)^2 \nonumber \\
{\cal M}^{(20)}_1(t-r^*) &=& {1\over 2} F_\chi^{(20)}(t-r^*)
- {1\over 14} F_\psi^{(v)}(t-r^*) F_\psi^{(iii)}(t-r^*)
- {3\over 14} F_Q^{(iv)}(t-r^*)^2 \nonumber \\
{\partial^2 {\cal M}^{(21)}_1(t-r^*)\over \partial t^2} &=&{1\over 2}
F_\chi^{(21)(ii)}(t-r^*) +{3\over 14} M F_\psi^{(v)}(t-r^*)^2 -
{12\over 7}F_\psi^{(iv)}(t-r^*)F_\psi^{(v)}(t-r^*) \nonumber \\ & &
-{3\over 14}F_\psi^{(ii)}(t-r^*) F_\psi^{(vii)}(t-r^*) - {9\over 14}
F_\psi^{(vi)}(t-r^*) F_\psi^{(iii)}(t-r^*)\nonumber \\ & & -{69\over
14} F_Q^{(v)}(t-r^*) F_Q^{(iv)}(t-r^*) +{9\over 14} F_Q^{(vi)}(t-r^*)
F_Q^{(iv)}(t-r^*) \nonumber \\ & & +{3\over 28} M F_\psi^{(v)}(t-r^*)
F_\psi^{(v)}(t-r^*) -{3\over 28} M F_\psi^{(iii)}(t-r^*)
F_\psi^{(vii)}(t-r^*) \nonumber \\ & & -{15\over 14} F_Q^{(vi)}(t-r^*)
F_Q^{(iii)}(t-r^*) +{9\over 14} M F_Q^{(v)}(t-r^*)^2 \nonumber \\
{\partial {\cal M}^{(20)}_2(t-r^*)\over \partial t} &=&F^{(20)}_Q(t-r^*)
-{1\over 14} F_Q^{(v)}(t-r^*) F_\psi^{(iii)}(t-r^*)
-{1\over 14} F_Q^{(iii)}(t-r^*) F_\psi^{(v)}(t-r^*)
\end{eqnarray}

We can now compute the gravitational waveforms by reading off the
appropriate metric components. For first order we find,
\begin{eqnarray}
{\partial G^{(1)}(r,t) \over \partial t} & = &{1\over r}
F_\psi^{(iv)}(t-r^*)+O(1/r^2) \nonumber \\ {\partial K^{(1)}(r,t)
\over \partial t} & = & {3\over r} F_\psi^{(iv)}(t-r^*)+O(1/r^2)
\nonumber \\ {\partial h^{(1)}_{2o}(r,t) \over \partial t} & = & - 2 r
F_Q^{(iv)}(t-r^*)+O(r^0) \label{wave1}
\end{eqnarray}
and the second order components are
\begin{eqnarray}
{\partial G^{(2)}(r,t) \over \partial t} & = & \left\{ \ F^{20}_\chi
(t-r^*)+ {1 \over 7}{\partial \over \partial t} \left
[ F_\psi^{(iv)}(t-r^*) F_\psi^{(iii)}(t-r^*) \right] -{3 \over 7}
F_Q^{(iv)}(t-r^*)^2 \right\}{1\over r} + O(1/r^2) \nonumber \\
{\partial K^{(2)}(r,t) \over \partial t} & = & 3 {\partial
G^{(2)}(r,t) \over \partial t} + O(1/r^2) \nonumber \\ {\partial
h^{(2)}_{2o}(r,t) \over \partial t} & = & -2 r \left\{ F^{20}_Q(t-r^*)
+ {1\over 7} {\partial^2 \over \partial t^2} \left[ F_Q^{(iii)}(t-r^*)
F_\psi^{(iii)}(t-r^*) \right] \right. \nonumber \\ & & \left. +{1
\over 7} F_Q^{(iv)}(t-r^*) F_\psi^{(iv)}(t-r^*) \right\} + O(r^0)
\end{eqnarray}
and in terms of first and second order Zerilli functions, we get
\begin{eqnarray}
{\partial G^{(1)}(r,t) \over \partial t} & = & {1\over r}
{\partial \psi^{(1)}\over \partial t}
+O(1/r^2)\nonumber\\ {\partial K^{(1)}(r,t) \over \partial
t} & = & 3 {\partial G^{(1)}(r,t) \over \partial t} \nonumber \\
{\partial h^{(1)}_{2o}(r,t) \over \partial t} & = & - 2 r
{\partial Q^{(1)}(r,t) \over \partial t}+O(r^0)
\end{eqnarray}
and 
\begin{eqnarray}
{\partial G^{(2)}(r,t) \over \partial t} & = & \left\{ \chi^{(2)}
+{1\over 7}{\partial \over \partial t} \left
[ {\partial\psi^{(1)}\over\partial t} \psi^{(1)} \right] - {3\over 7}
\left[ {\partial Q^{(1)}\over\partial t} \right]^2 \right\}{1\over r}
+O(1/r^2) \nonumber \\ {\partial K^{(2)}(r,t) \over \partial t} & = &
3 {\partial G^{(2)}(r,t) \over \partial t} + O(1/r^2) \nonumber \\
{\partial h^{(2)}_{2o}(r,t) \over \partial t} & = &-2 r \left
\{ \Theta^{(2)} +{1\over 7} {\partial^2\over\partial t^2} \left[ \psi^{(1)}
Q^{(1)} \right] +{1\over 7} {\partial\over\partial t} \psi^{(1)}
{\partial\over\partial t} Q^{(1)} \right\} + O(r^0)
\end{eqnarray}

It is straightforward to compute the radiated energy, since we are in
an explicitly asymptotically flat gauge. One can therefore simply
apply the formulae stemming from Landau and Lifshitz \cite{LaLi,CPM1},
\begin{equation}
{d {\rm Power}\over d\Omega} = {1 \over 16\pi r^2} 
\left[\left({\partial\over \partial t} h_{\theta\phi}\right)^2+
{1 \over 4} \left({\partial \over \partial t} h_{\theta\theta}
-{1 \over \sin^2\theta}{\partial \over \partial t}h_{\phi\phi}\right)^2\right].
\end{equation}
For the perturbations we are considering,
\begin{eqnarray}
{h}_{\theta\theta}&=& r^2 \left\{[\epsilon
\widetilde{K}^{(1)}(r,t)+\epsilon^2 \widetilde{K}^{(2)}(r,t)]
P_2(\theta) \right. \nonumber \\ & & \left.+ [\epsilon
\widetilde{G}^{(1)}(r,t) + \epsilon^2 \widetilde{G}^{(2)}(r,t)]
\partial^2 P_2(\theta) /\partial \theta^2)\right\}\\
{h}_{\phi\phi} &=&r^2 \left\{ \sin^2\theta [\epsilon
\widetilde{K}^{(1)}(r,t) + \epsilon^2 \widetilde{K}^{(2)}(r,t)]
P_2(\theta) \right. \nonumber \\ & & \left.+\sin(\theta) \cos(\theta)[
\epsilon \widetilde{G}^{(1)}(r,t) +\epsilon^2
\widetilde{G}^{(2)}(r,t)] \partial P_2(\theta) /\partial
\theta)\right\}\\
{h}_{\theta \phi}&=& {1 \over 2} \left[\epsilon
\widetilde{h}^{(1)}_{2o}(r,t)+ \epsilon^2
\widetilde{h}^{(2)}_{2o}(r,t)\right] \left[ \cos(\theta) \partial
P_2(\theta)/\partial\theta- \sin(\theta) \partial^2
P_2(\theta)/\partial\theta^2 \right] 
\end{eqnarray}
and substituting, we get,
\begin{eqnarray}
{\rm Power} &=& {3\over 10}\left[\epsilon {\partial \psi^{(1)} \over
\partial t} +\epsilon^2 \left(\chi^{(2)}+{1 \over 7} {\partial \over
\partial t} \left(\psi^{(1)} {\partial \psi^{(1)} \over \partial
t}\right) - {3\over 7}\left({\partial Q^{(1)}\over \partial
t}\right)^2\right)\right]^2 \nonumber\\ 
&+& {36\over 35} \left[\epsilon
{\partial Q^{(1)}\over 
\partial t}+\epsilon^2 \left(\Theta^{(2)} + {1 \over 7}\left({\partial^2
Q^{(1)}\psi^{(1)} \over \partial t^2} + {\partial Q^{(1)}\over \partial
t} {\partial \psi^{(1)} \over \partial t}\right)\right)\right]^2.
\end{eqnarray}
As can be seen, the energy has two quadratic contributions, one coming
from the $h_{\theta\phi}$ component of the metric and one from the
diagonal elements of it. They correspond to the ``$\times$'' and
``$+$'' polarization modes of the gravitational field and therefore we
can consider the expressions in the square brackets as
the``waveforms'' of the gravitational waves emitted. Contrary to the
first order case, in which the waveform is directly given by the
Zerilli (or Regge--Wheeler) function, for second order corrections the
relationship is more involved.

\section{Initial data}

Initial data for solutions of Einstein equations is usually given in
terms of the initial value of the three-metric and the initial value
of the extrinsic curvature of the initial Cauchy surface. For the case
of initial data for colliding black holes, the popular families of
initial data of Bowen and York \cite{BoYo} and punctures \cite{BrBr}
naturally come cast in a gauge in which\footnote{ This happens to
coincide exactly with the same gauge condition of the asymptotically
flat gauge.}  $H_0=0$, $ H_1=0$, $ h_0=0$ and $ h_{0o}=0$. We will
therefore present the appropriate formulae that would allow us, given
a metric and extrinsic curvature cast in such a gauge, to provide the
initial data for the first and second order Zerilli functions and
their time derivatives.

The extrinsic curvature $K_{ab}$ of a spatial slice is defined by
\begin{equation}
K_{ab}\,=\,n_{(c;d)}\,\left( {\delta^c}_{a} + n^c n_a \right) \left(
{\delta^d}_{b} + n^d n_b \right)
\end{equation}
where $n_a$ is the unit normal to the Cauchy surface where the data
are specified.  One can relate the extrinsic curvature to the time 
derivative of the three-metric through,
\begin{eqnarray}
\label{ADMeqs}
ds^2 & = & g_{ij} dx^i dx^j - N^2 (dt)^2 \nonumber \\
{\partial g_{ij} \over \partial t} & = & -2NK_{ij}.
\end{eqnarray}
That is, if through some procedure we are given the extrinsic
curvature and the three metric, we can compute the time derivative of
the three metric. Re-expressing this in terms of components of the
tensor spherical harmonic decomposition due to Regge and Wheeler 
we get,
\begin{equation}
\begin{array}{rclrcl}
{\partial h^{(1)}_1 / \partial t} & = & - 2 \sqrt{1-2 M/r} \,
K^{(1)}_{(h_1)} & {\partial H^{(1)}_2 / \partial t} & = & - 2
\sqrt{1-2 M/r} \, K^{(1)}_{(H_2)} \nonumber \\ {\partial K^{(1)} /
\partial t} & = & - 2 \sqrt{1-2 M/r} \, K^{(1)}_{(K)} & {\partial
G^{(1)} / \partial t} & = & - 2 \sqrt{1-2 M/r} \, K^{(1)}_{(G)}
\nonumber \\ {\partial h^{(1)}_{1o} / \partial t} & = & - 2 \sqrt{1-2
M/r} \, K^{(1)}_{(h_{1o})} & {\partial h^{(1)}_{2o} / \partial t} & =
& - 2 \sqrt{1-2 M/r} \, K^{(1)}_{(h_{2o})} \nonumber \\
\end{array}
\end{equation}
for first order components. $K^{(1)}_{(xx)}$ are spherical tensor 
harmonic components of the first order extrinsic curvature tensor, 
written in the Regge--Wheeler notation that we used in formulae
(\ref{rw1}-\ref{rwe}). For second order we get,
\begin{equation}
\begin{array}{rclrcl}
{\partial h^{(2)}_1 / \partial t} & = & - 2 \sqrt{1-2 M/r} \,
K^{(2)}_{(h_1)} & {\partial H^{(2)}_2 / \partial t} & = & - 2
\sqrt{1-2 M/r} \, K^{(2)}_{(H_2)} \nonumber \\ {\partial K^{(2)} /
\partial t} & = & - 2 \sqrt{1-2 M/r} \, K^{(2)}_{(K)} & {\partial
G^{(2)} / \partial t} & = & - 2 \sqrt{1-2 M/r} \, K^{(2)}_{(G)}
\nonumber \\ {\partial h^{(2)}_{1o} / \partial t} & = & - 2 \sqrt{1-2
M/r} \, K^{(2)}_{(h_{1o})} & {\partial h^{(2)}_{2o} / \partial t} & =
& - 2 \sqrt{1-2 M/r} \, K^{(2)}_{(h_{2o})} \nonumber \\
\end{array}
\end{equation}
Here $K^{(2)}_{(xx)}$ are spherical harmonic components of the second
order extrinsic curvature tensor. Note that the expressions are
similar to the first order ones only in the gauge we are considering. 
In other gauges (in which the perturbations of lapse and shift may not
vanish), there will be extra terms, quadratic in first order elements.

A case in which the formalism just developed will be of use (and we
will discuss in the forthcoming publication) is that of the collision
of counterrotating Bowen--York holes (the cosmic screw case). In such
a case, to first order the metric has even portions only (similar to
those in the Misner \cite{Mi} initial data) and the extrinsic
curvature odd portions. So putting together the above formulae, for 
this case one gets,
\begin{eqnarray}
\chi^{(2)}&=& -{2\over 7} { \sqrt{r-2 M} \over (2 r+3 M) \sqrt{r} }
\left[ 7\,{r}^{2} K^{(2)}_{(K)} -7\,{r}^{2} K^{(2)}_{(G)} -7\,
r K^{(2)}_{(h_1)} +2\,\left ( K^{(1)} \right )^{2}r
\sqrt {-{\frac {r}{-r+2\,M}}}+ \right. \nonumber \\
&+&\left. 3\,\left ( K^{(1)} 
\right )^{2}M\sqrt {-{\frac {r}{-r+2\,M}}}
+14\, K^{(2)}_{(h_1)} M+21\,r K^{(2)}_{(G)} M \right] \nonumber \\
{\partial \chi^{(2)} \over \partial t} &=&-{1\over 14} {r -2 M\over
r^4 (2 r +3 M)} \left[ 140\,Mr h_1^{(2)} +84\,{M}^{2}{r}^{2} G^{(2)}_r
+{r}^{5 }\left ( K^{(1)}_r \right )^{2}-15\,{r}^{3}\left ( K^{(1)}
\right )^{2}-2\,M{r}^{4}\left ( K^{(1)}_r \right )^{2}+
\right. \nonumber \\ &+& \left. 2\,M{r}^{2}\left ( K^{(1)} \right
)^{2}+ 2\,{r}^{4} K^{(1)} K^{(1)}_r +36\,{r}^{3}\left
( K^{(1)}_{(h_{1o})} \right )^{2}+96\,r\left ( K^{(1)}_{(h_{2o})}
\right )^{2}+276\,M\left ( K^{(1)}_{(h_{2o})} \right
)^{2}-\right. \nonumber \\ &-& \left. 144\,{M}^{2}r\left
( K^{(1)}_{(h_{1o})} \right )^{2}+72\,{r}^{2} K^{(1)}_{(h_{1o})}
K^{(1)}_{(h_{2o})} +12\,{r}^{2} K^{(1)}_{(h_{2o})} K^{(1)}_{(h_{2o})r}
-14\,M{r}^{3} K^{(2)}_r +28\,{r}^{3} K^{(2)} - \right. \nonumber \\
&-& \left. 28\,{r}^{5} G^{(2)}_{rr} -28\,{r}^{4} G^{(2)}_r -144\,Mr
K^{(1)}_{(h_{1o})} K^{(1)}_{(h_{2o})} -24\,Mr K^{(1)}_{(h_{2o})}
K^{(1)}_{(h_{2o})r} + \right. \nonumber \\ &+& \left. 84\,{M}^{2}{r
}^{3} G^{(2)}_{rr} +14\,M{r}^{4} G^{(2)}_{rr} -42\,M{r} ^{3} G^{(2)}_r
-168\,{M}^{2}r h_{1\;r}^{(2)} -28\,M{r}^ {2} h_{1\;r}^{(2)}
+56\,{r}^{3} h_{1\;r}^{(2)} + \right. \nonumber \\ &+& \left. 168\,{M
}^{2} h_1^{(2)} -56\,{r}^{2} h_1^{(2)} -28\,{r}^{3} H_2^{(2)}
-28\,M{r}^{2} H_2^{(2)} \right] \nonumber \\
\Theta^{(2)}&=& -{\frac {\left (-r+2\,M\right )\left (-2\,
 h_{2o}^{(2)} +r h_{2o\;r}^{(2)} +2\,r h_{1o}^{(2)} 
\right )}{2\,{r}^{3}}} \nonumber \\
{\partial \Theta^{(2)}\over \partial t}&=& {1\over 7} {\sqrt{r-2 M}\over
r^3 \sqrt{r}} \left[ {r}^{2} K^{(1)}_{(h_{1o})} K^{(1)} -14\,{r}^{2}
K^{(2)}_{(h_{1o})} -7\,{r}^{2} K^{(2)}_{(h_{2o})r} -35 \,
K^{(2)}_{(h_{2o})} M+14\,r K^{(2)}_{(h_{2o})} + \right. \nonumber \\
&+& \left. 6\,r K^{(1)}_r K^{(1)}_{(h_{2o})} M+8\,r K^{(1)}
K^{(1)}_{(h_{2o})r} M+14\,r K^{(2)}_{(h_{2o})r} M+28\, r
K^{(2)}_{(h_{1o})} M-4\,{r}^{2} K^{(1)} K^{(1)}_{(h_{2o})r} -
\right. \nonumber \\ &-& \left. 3\,{r}^{2} K^{(1)}_r
K^{(1)}_{(h_{2o})} +8\,r K^{(1)} K^{(1)}_{(h_{2o})} - 20\, K^{(1)}
K^{(1)}_{(h_{2o})} M-2\,r K^{(1)}_{(h_{1o})} K^{(1)} M \right]
\end{eqnarray}

\section{Conclusions}

We have written explicitly the second order perturbative equations for
axisymmetric $\ell=2$ perturbations including odd parity modes. We set
up explicitly the evolution equations and the formulae for generating
initial data starting from a metric and an extrinsic curvature. We
also performed an asymptotic analysis to obtain formulas for radiated
energies and waveforms. These formulae will be useful for considering
perturbatively the evolution of spacetimes arising from the collision
of nearby black holes. They might also be useful for other comparison
with nonlinear evolutions, as the ones considered in
\cite{Baker:1999}.

\acknowledgements

This work was supported in part by grants
of the National Science Foundation of the US INT-9512894, PHY-9423950,
PHY-9800973, PHY-9407194, by funds of the University of C\'ordoba, the
Pennsylvania State University and the Eberly Family Research Fund at
Penn State. We also acknowledge support of CONICET and CONICOR
(Argentina). JP also acknowledges support from the Alfred P. Sloan and
John S. Guggenheim Foundations.  RJG is a member of CONICET.

\end{document}